\documentstyle[12pt,epsfig,here,procsla]{article}
  
  \catcode`\@=11
  \long\def\@makefntext#1{
  \protect\noindent \hbox to 3.2pt {\hskip-.9pt  
  $^{{\ninerm\@thefnmark}}$\hfil}#1\hfill}		
  
  \def\@makefnmark{\hbox to 0pt{$^{\@thefnmark}$\hss}}  
  	
  \def\ps@myheadings{\let\@mkboth\@gobbletwo
  \def\@oddhead{\hbox{}
  \rightmark\hfil\ninerm\thepage}   
  \def\@oddfoot{}\def\@evenhead{\ninerm\thepage\hfil
  \leftmark\hbox{}}\def\@evenfoot{}
  \def\sectionmark##1{}\def\subsectionmark##1{}}
  
\begin{document}

\vspace*{3mm}  
  \centerline{\Large \bf AMANDA: status, results and future}
  \baselineskip=16pt
  
\vspace*{0.6cm}
  \centerline{\footnotesize presented by Christian Spiering for the
   AMANDA collaboration:}

\bigskip
{\footnotesize
\noindent
E.~Andres$^{11}$, 
P.~Askebjer$^{4}$, 
G.~Barouch$^{8}$, 
S.~Barwick$^{6}$, 
X.~Bai$^{11}$, 
K.~Becker$^{9}$, 
R.~Bay$^{5}$, 
L.~Bergstr\"om$^{4}$, 
D.~Bertrand$^{12}$, 
D.~Besson$^{13}$, 
A.~Biron$^{2}$, 
J.~Booth$^{6}$, 
O.~Botner$^{14}$, 
A.~Bouchta$^{2}$, 
S.~Carius$^{3}$, 
M.~Carlson$^{8}$, 
W.~Chinowsky$^{10}$, 
D.~Chirkin$^{5}$, 
J.~Conrad$^{14}$, 
C.~Costa$^{8}$, 
D.~F.~Cowen$^{7}$, 
E.~Dalberg$^{4}$, 
J.~Dewulf$^{12}$, 
T.~DeYoung$^{8}$, 
J.~Edsj\"o$^{4}$, 
P.~Ekstr\"om$^{4}$, 
G.~Frichter$^{13}$, 
A.~Goobar$^{4}$, 
L.~Gray$^{8}$, 
A.~Hallgren$^{14}$, 
F.~Halzen$^{8}$, 
Y.~He$^{5}$, 
R.~Hardtke$^{8}$, 
G.~Hill$^{8}$, 
P.~O.~Hulth$^{4}$, 
S.~Hundertmark$^{2}$, 
J.~Jacobsen$^{10}$, 
V.~Kandhadai$^{8}$, 
A.~Karle$^{8}$, 
J.~Kim$^{6}$, 
B.~Koci$^{8}$, 
M.~Kowalski$^{2}$, 
I.~Kravchenko$^{13}$, 
J.~Lamoureux$^{10}$, 
P.~Loaiza$^{14}$, 
H.~Leich$^{2}$, 
P.~Lindahl$^{3}$, 
T.~Liss$^{5}$, 
I.~Liubarsky$^{8}$, 
M.~Leuthold$^{2}$, 
D.~M.~Lowder$^{5}$, 
J.~Ludvig$^{10}$, 
P.~Marciniewski$^{14}$, 
T.~Miller$^{1}$, 
P.~Miocinovic$^{5}$, 
P.~Mock$^{6}$, 
F.~M.~Newcomer$^{7}$, 
R.~Morse$^{8}$, 
P.~Niessen$^{2}$, 
D.~Nygren$^{10}$, 
C.~P\'erez~de~los~Heros$^{14}$, 
R.~Porrata$^{6}$, 
P.~B.~Price$^{5}$, 
G.~Przybylski$^{10}$, 
K.~Rawlins$^{8}$, 
W.~Rhode$^{5}$, 
S.~Richter$^{11}$, 
J.~Rodriguez~Martino$^{4}$, 
P.~Romenesko$^{8}$, 
D.~Ross$^{6}$, 
H.~Rubinstein$^{4}$, 
E.~Schneider$^{6}$, 
T.~Schmidt$^{2}$, 
R.~Schwarz$^{11}$, 
A.~Silvestri$^{2}$, 
G.~Smoot$^{10}$, 
M.~Solarz$^{5}$, 
G.~Spiczak$^{1}$, 
C.~Spiering$^{2}$, 
N.~Starinski$^{11}$, 
P.~Steffen$^{2}$, 
R.~Stokstad$^{10}$, 
O.~Streicher$^{2}$, 
I.~Taboada$^{7}$, 
T.~Thon$^{2}$, 
S.~Tilav$^{8}$, 
M.~Vander~Donckt$^{12}$, 
C.~Walck$^{4}$, 
C.~Wiebusch$^{2}$, 
R.~Wischnewski$^{2}$, 
K.~Woschnagg$^{5}$, 
W.~Wu$^{6}$, 
G.~Yodh$^{6}$, 
S.~Young$^{6}$

\noindent
1) Bartol Research Institute, University of Delaware, Newark, DE, USA \\
2) DESY-Zeuthen, Zeuthen, Germany \\
3) Kalmar University, Sweden \\
4) Stockholm University, Stockholm, Sweden \\
5) University of California, Berkeley, Berkeley, CA, USA \\
6) University of California,  Irvine, Irvine, CA, USA \\
7) University of Pennsylvania, Philadelphia, PA, USA \\
8) University of Wisconsin, Madison, WI, USA \\
9) University of Wuppertal, Wuppertal, Germany \\
10) Lawrence Berkeley Laboratory, Berkeley, CA, USA \\
11) South Pole Station, Antarctica \\
12) University of Brussels, Brussels, Belgium \\
13) University of Kansas, Lawrence, KS, USA \\
14) University of Uppsala, Uppsala, Sweden 
}

\vspace{0.9cm}
\abstracts{We review the status of the AMANDA neutrino telescope. 
We present results obtained from the four-string prototype
array
AMANDA-B4 and describe the methods of track reconstruction and
neutrino event separation. We give also first results
of the analysis of the 10-string detector AMANDA-B10, in particular on
atmospheric neutrinos and the search for magnetic monopoles.
We sketch the future schedule on the
way to a cube kilometer telescope at the South Pole, ICECUBE.}

\newpage
   
\section{The Detector}

\noindent
AMANDA (Antarctic Muon And Neutrino Detector Array) uses the natural
Antarctic ice as both target and Cherenkov medium \cite{Lowder,Amanda}. 
The detector consists of  
strings of optical modules (OMs) frozen in the 
3 km thick ice sheet at the South Pole. An OM consists of an
$8^{\prime \prime}$
photomultiplier in a  glass vessel. The strings are 
deployed into holes drilled with pressurized hot water. The 
water column in the hole then refreezes within 35-40 hours, fixing 
the string in its final position. In our basic design, each OM had 
its own cable supplying the high voltage (HV)
as well as transmitting the anode signal. For the last
122 OMs deployed in the antarctic season 1998/99, the anode
signal drives a LED which's signal is transmitted via an optical
fiber. Other approaches to signal transmission
are described in section 6.

\begin{figure}[htbp]
\centering
\hspace{0.5cm}
\mbox{\epsfig{file=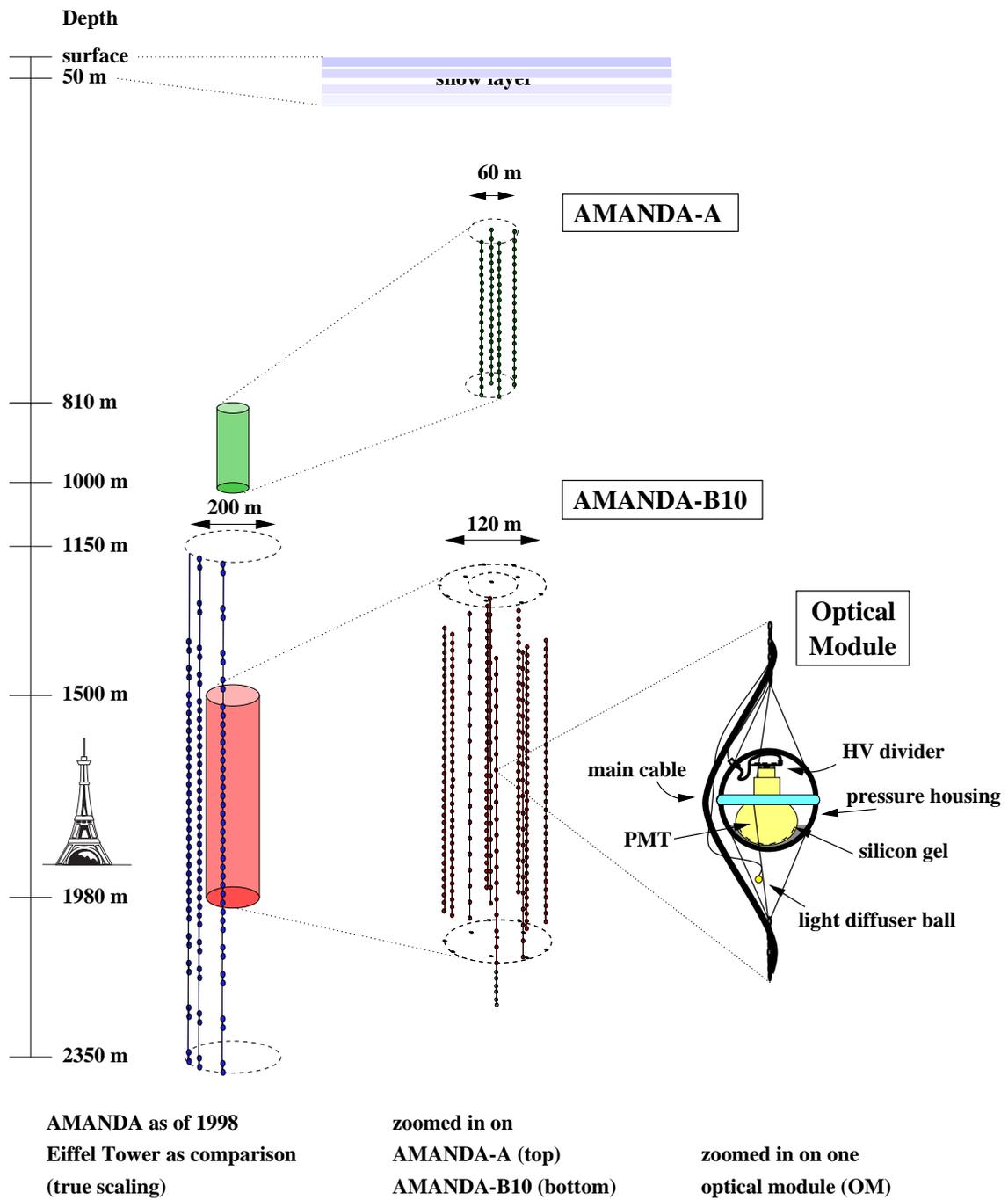,height=18.0cm}}
              \caption {\small \label{fullamanda}
      Scheme of the 1998 AMANDA installations. The left picture  is
drawn with true scaling. A zoomed view on AMANDA-A
(top) and AMANDA-B10 (bottom) is shown at the center. The right zoom 
depicts the optical module.
      }
\end{figure}

Fig.~\ref{fullamanda} shows the current configuration of 
the AMANDA detector.
The shallow array, AMANDA-A, was deployed at a depth 
of 800 to 1000\,m
in 1993/94 in an exploratory phase of the project.
Studies of the optical properties of the ice  
carried out with AMANDA-A showed that a high 
concentration of residual air bubbles remaining at these depths  
leads to strong scattering of light, making 
accurate track reconstruction impossible. 
Therefore, in the polar season
1995/96 a deeper array consisting of 86 OMs arranged on four strings
(AMANDA-B4) 
was deployed at depths ranging from 1540 to 2040 meters, where the 
concentration of bubbles was predicted to be negligible according to  
extrapolation of AMANDA-A results. 
The detector was upgraded in 1996/97 with 216 additional OMs on
6 strings. This detector of 4+6 strings was named AMANDA-B10 and is sketched
at the right side of fig.~\ref{fullamanda}. AMANDA-B10 was upgraded in the 
season 1997/98 by 3 strings instrumented between 1150~m and 1350~m
which fulfill several
tasks. Firstly, they  explore the very deep and very shallow ice
with respect to a future cube kilometer array. Secondly, they
form one corner of AMANDA-II which is the next stage of AMANDA
with altogether about 700 OMs. Thirdly, they have been used
to test new technologies of data transmission.

An essential ingredient to the operation of a detector like 
AMANDA is the  knowledge of the optical properties of the 
ice, as well as a precise geometry and time calibration of the detector. 
We make use of the following calibration tools:
Pulsed light sources are used to determine {\it a)}
time offsets, {\it b)} the geometry of the array, and {\it c)}
to derive ice properties. They include
a YAG laser calibration system which transmits light pulses
from a YAG laser at the surface via optical fibers 
to diffuser balls located at each PMT, as well as  
nitrogen lasers and LED beacons at various depths.
DC light sources allow to measure the attenuation of light.
Another calibration source are muons themselves. The response of the array
to muons allows to derive time offsets and ice properties in a
way alternative to that using dedicated light sources.
Finally, drill recording and pressure sensors give the absolute
positions of the strings.
Since values obtained for time offsets, geometry and ice properties 
are dependent on each other, the calibration process is 
non-trivial and time consuming.
After having worked through  appropriate procedures,
the B10 time offsets are now known with about 5~nsec accuracy
(which is comparable to the 1 photoelectron time jitter of about 4 nsec), 
and the relative positions of OMs with an accuracy of 0.5-1.0~m.

\begin{center}
\begin{figure}[htbp]
\centering
\hspace{0.5cm}
\mbox{\epsfig{file=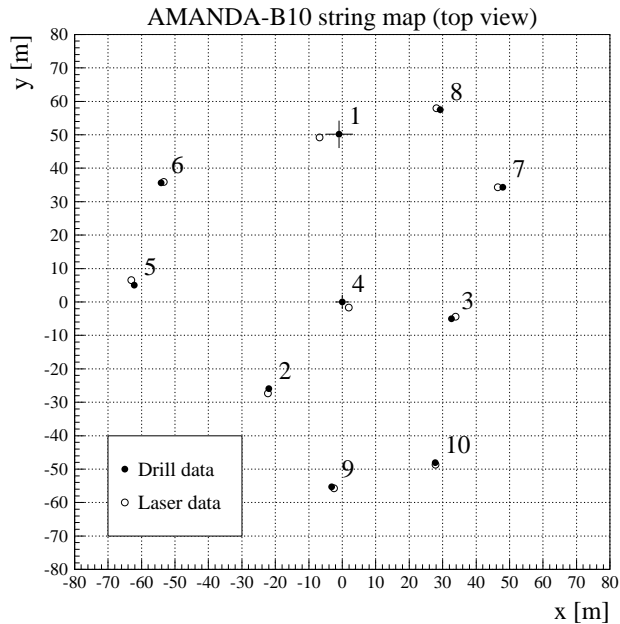,height=8.9cm}}
\vspace{-3mm}
              \caption {\small
Top view of AMANDA-B10. Open circles denote the positions
determined by the laser system, closed circles denote
positions obtained from drilling information.
}
\label{Kurt}
\end{figure}
\end{center}

Fig.\ref{Kurt} shows the top view of the B10 detector, with
the open circles giving the results of the laser calibration, and
the filled circles the results of the drill logging data.
With the exception of strings 1 and 4 (which are slightly tilted 
and cannot be handled exactly by the laser analysis) 
one observes agreement within 1 meter.

\vspace{-5mm}

\begin{center}
\begin{figure}[htbp]
\centering
\mbox{\epsfig{file=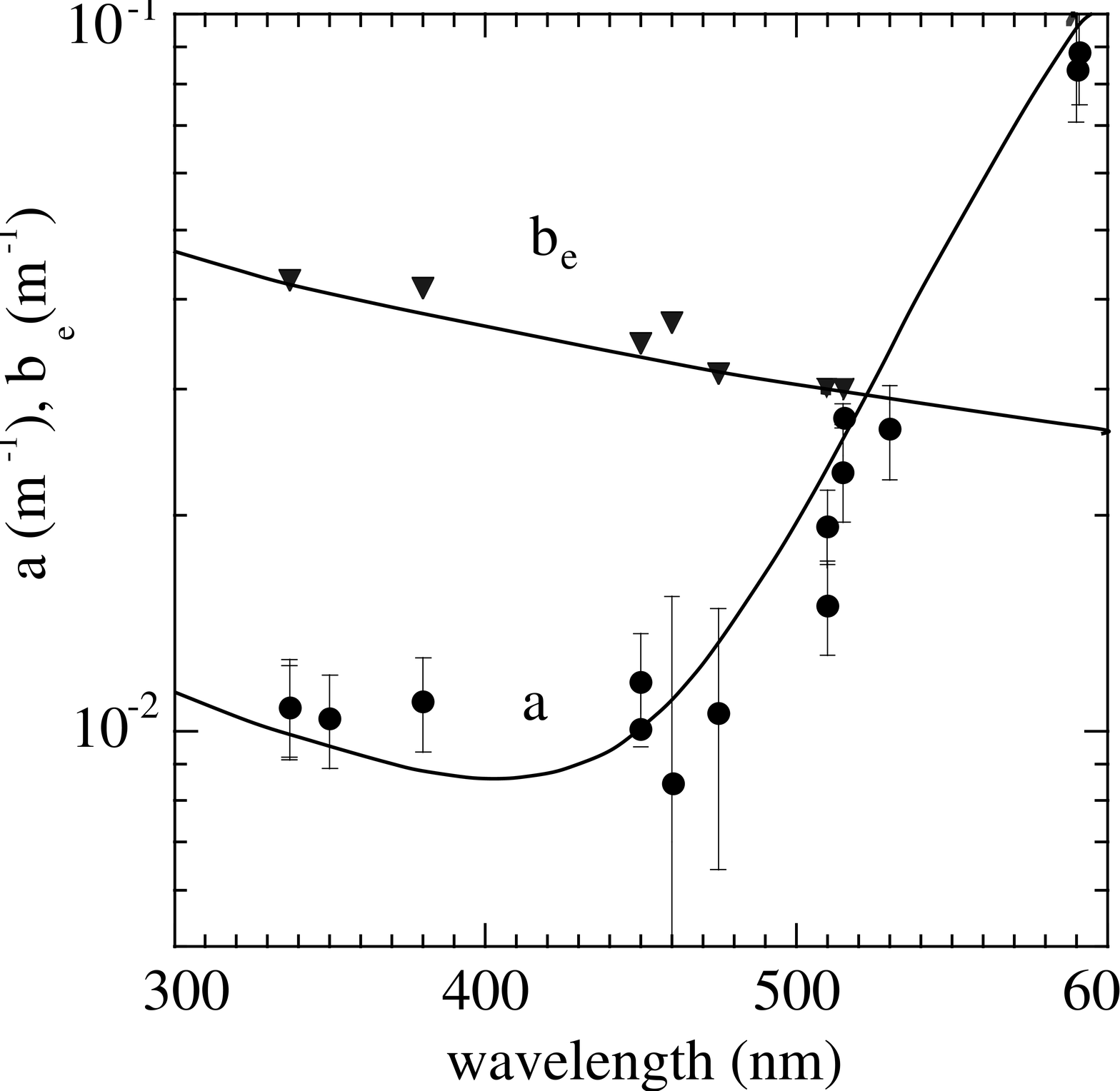,height=8.7cm}}
\caption {\small
Absorption and scattering coefficients at an average depth
of 1.7~km.
}
\label{Buford}
\end{figure}
\end{center}

\vspace{-6mm}
Fig.\ref{Buford} shows data on the wavelength dependence of scattering 
and absorption compared to theory of He and Price 
\cite{Yudong}. The absorption length $1/a$ is between 90 and 100~m
for wavelengths below 460~nm, i.e. ice absorbs not only about half 
as much as ocean water, but also does not degrade in
transparency towards smaller wavelengths down to 337 nm. 
On the other hand, scattering is nearly an order of
magnitude
stronger than in water: the effective scattering length 
$1/b =$  geometric scattering 
length/$(1 -  \langle \cos{\theta} \rangle)$ varies
between 24 and 30 m in the relevant wavelength range.
$\langle \cos \theta \rangle $ is the average cosine of the
scattering angle and is supposed to be about 0.8 in deep ice.
These values vary with depth by $\pm 30\%$ between 1.5 and
2.0~km \cite{Price}.

\vspace*{-12mm}
\section{Reconstruction of Muon Tracks}
\noindent
The reconstruction procedure for a muon track consists of
five steps:

\noindent
1. Rejection of noise hits.

\noindent
2. A line approximation \cite{Stenger} which yields a 
first track estimate and a velocity  $\vec{v}$.

\noindent
3. A likelihood fit based on the measured times.
         This "time fit" yields angles and coordinates
        of the track as well as a likelihood ${\cal L}_{time}$.

\noindent
4. A likelihood fit using the fitted track parameters from the time fit
         and varying the light emission per unit length until the probabilities
         of the hit PMTs to be hit and non-hit PMTs to be not hit are
         maximized. This fit does not vary the direction of the track
         but yields a likelihood ${\cal L}_{hit}$
         with can be used as a quality parameter.
         
\noindent
5. A quality analysis applying cuts in order to reject badly 
          reconstructed events.

\subsection{Time Fit}

\begin{figure}[htbp]
\begin{center} 
\mbox{\epsfig{file=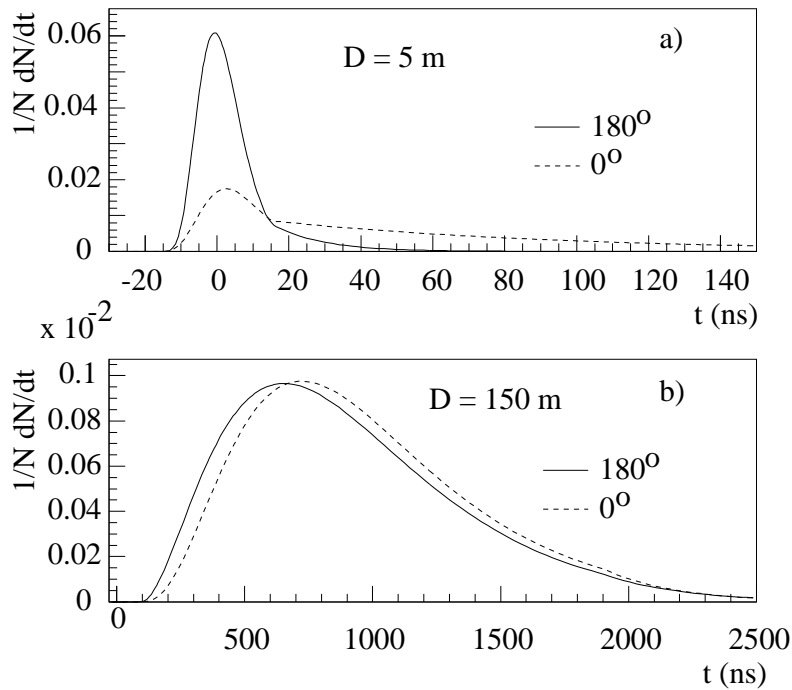,height=9cm}} 
\caption[2]{\small
Arrival time distributions for modules facing (full curves) and 
back-facing (dashed curves) a muon track. Parameterizations 
\cite{bouchta} are shown for muon tracks with impact 
parameters of 5 meters (a) and 150 meters (b).
}
\label{adam}
\end{center}
\end{figure}

\noindent
In an ideal medium without scattering, one would reconstruct
the path of  minimum ionizing muons most efficiently 
by a $\chi^2$ minimization.
Because of scattering in ice, the distribution of arrival times
of photoelectrons seen by a PMT is not Gaussian
but has a long tail at the high side -- see fig.~\ref{adam}. 
In order to cope with the non-Gaussian timing distributions
we used a likelihood analysis. In this approach, a
normalized probability distribution function $p_i(t)$  gives the 
probability of a certain time delay $t$ for a given hit $i$
with respect to straightly propagating photons. 
This probability function is derived from  MC simulations
of photon propagation in ice.
By varying the  track parameters the logarithm of a 
likelihood function ${\cal L}$  is maximized.

$$
\log ({\cal L}) = \log \left ( \prod_{\mbox{all hits}} p_i \right )
= \sum_{\mbox{all hits}} \log ( p_i ) 
$$

\noindent
In order to be used in the iteration process, the time delays  as obtained
from the photon propagation Monte-Carlo have to be parameterized
by an analytic formula.
The AMANDA collaboration has developed two independent reconstruction
programs, using different parameterizations of
the photon propagation as well as different minimization methods 
\cite{bouchta,icrc_reco}.
Both methods are in good agreement with each other.
Fig.~\ref{adam} shows the result of the parameterization of the time
delay for two distances and for two angles between the PMT axis
and the muon direction.

At a distance of 5\,m and a PMT facing toward the muon
track, the delay curve is dominated by the time jitter of the
PMT. If the PMT looks into the opposite direction,
the contribution of scattered photons yields a long
tail towards large delays. At distances as large as 150\,m,
distributions for both directions of the PMT are close to each
other since all photons reaching the PMT are multiply scattered.

\subsection{Quality Analysis}

\noindent
Quality criteria are applied in order to
select events which are "well" reconstructed.  
The criteria define cuts on topological event parameters
and observables derived from the 
reconstruction, e.g.
\vspace{-2mm}
\begin{itemize}
\item 
Speed $|\vec{v}|$ of the line approximation. 
Values close to the speed of light
indicate a reasonable pattern of the measured times.
\vspace{-2mm}
\item
"Time" likelihood per hit PMT $\log({\cal L}_{time})/N_{hit}$. 
\vspace{-2mm}
\item 
"Hit" likelihood per all working channels, 
$\log({\cal L}_{hit})/N_{all}$.
\vspace{-2mm}
\item 
Number of direct hits, $N_{dir}$, which is defined to be
the number of hits with time residuals $t_i\mbox{(measured)} - 
t_i\mbox{(fit)}$
smaller than a certain cut value. We use cut values of 15\,nsec,
25\,nsec and 75\,nsec, and denote the corresponding
parameters as $N_{dir}$(15), $N_{dir}$(25) and
$N_{dir}$(75), respectively. 
Events with more than a certain minimum number of direct
hits (i.e. only slightly delayed photons) are likely to be
well reconstructed \cite{icrc_reco}.
\vspace{-2mm}
\item
The projected length of direct hits onto the reconstructed 
track, $L_{dir}$.
A cut in this parameter rejects events with a small lever arm.
\vspace{-2mm}
\item
Vertical coordinate of the center of gravity, $z_{COG}$.
Cuts on this parameter are used to reject events close
to the borders of the array. 
\end{itemize}

\section{Verification of reconstruction results by SPASE coincidences}

\noindent
AMANDA is unique in that it can be
calibrated by muons with
known zenith and azimuth angles which are tagged by 
air shower detectors at the surface. AMANDA-B4 has been running in
coincidence with the two SPASE (South Pole Air Shower Experiment)
arrays, SPASE-1 
and SPASE-2 \cite{Gais95} and the GASP Air Cherenkov Telescope. 
SPASE-2 is located
370\,m away from the center of AMANDA.
It consists of 30 scintillator stations
on a 30\,m triangular grid, with a total area of
$1.6 \cdot 10^4$\,m$^2$.
For each air shower, the direction,
core location, shower size and GPS time are determined. 

\begin{figure}[htbp]
\centering
\mbox{\epsfig{file=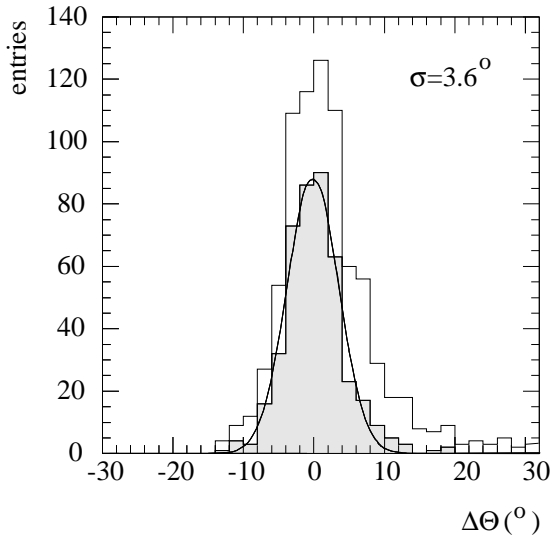,height=7cm}}
\caption [11]{\small
Mismatch between zenith angles determined in AMANDA-B4 and SPASE-2.
}
\label{Spase2}
\end{figure}

A one-week sample 
of SPASE-2--AMANDA coincidences  has been analyzed in order to compare
the directions of muons determined by AMANDA-B4 to those
of the showers measured by SPASE-2. A histogram of the zenith mismatch 
angle between SPASE-2 and AMANDA-B4 
is  shown in fig.\ref{Spase2}. 
The selected events are required to 
have $\ge$8 hits along 3 strings
and to yield a track which is closer than 150\,m to the
air shower axis measured by SPASE-2 (upper histogram).
The hatched histogram shows the distribution of the zenith
mismatch angle after requesting
$\log({\cal L}_{time})/N_{hit} > $ -12,
$N_{dir}\mbox{(75)} > 4$ and 
$L_{dir}\mbox{(75)} > 50$\,m. 

428 of the originally 
840 selected events pass these
quality cuts. The gaussian fit has a mean of
$(0.14 \pm 0.19)$ degrees and a width of $\sigma = (3.6 \pm 0.17)$ 
degrees. 
The small mean implies that there is little 
systematic error in zenith angle
reconstruction.
The SPASE-2 pointing accuracy
depends on zenith angle and shower size
and is typically
between 1$^\circ$ and 2$^\circ$ \cite{Spase2}.

\section{Results from AMANDA-B4}

\vspace{-5mm}
\subsection{Intensity-vs-Depth Relation for Atmospheric Muons \label{depth}}

\noindent
The muon intensity $I(\theta_{\mu})$ as a function of the zenith angle 
$\theta$ is obtained from 

\begin{equation}\label{fluxform1}
   I(\theta_{\mu})=   \frac{S_{dead}}{T \cdot \Delta \Omega}\, 
   \frac{ 
   N_{\mu}(\theta_{\mu})\,
   \cdot m(\theta_{\mu})}
   {{\epsilon_{rec}(\theta_{\mu})} \cdot
   {A_{eff}}(\mbox{cut},\theta_{\mu})}
\end{equation}

\noindent
$N_{\mu}(\theta_{\mu}$) is
the number of events with a reconstructed zenith angle 
$\cos(\theta_{\mu})$. $T=22.03$ hours is the data time used for the
atmospheric muon analysis.
$S_{dead}$=1.14 accounts for the deadtime of 
the data acquisition.
$\Delta\Omega$ is the solid angle covered by the corresponding 
$\cos(\theta_{\mu})$ interval. 
$A_{eff}(\mbox{cut},\theta_{\mu})$ is the effective area at zenith 
angle $\theta_{\mu}$. 
The reconstruction efficiency 
$\epsilon_{rec}(\theta_{\mu})$ is typically 0.8.
The mean muon multiplicity $m(\theta_{\mu})$ is about 1.2 for vertical
tracks and decreases towards the horizon.

\begin{figure}[htb]
\vspace{-1.cm}
\centering
\mbox{\epsfig{file=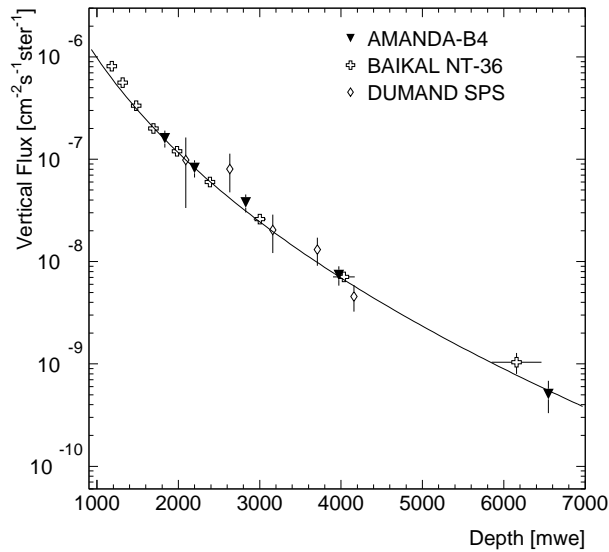,width=8cm}}
\caption[2]{\small \label{ABD}
Vertical intensity versus depth for Amanda, Baikal and
Dumand. The full line gives the prediction of
\cite{bugaev}.
}
\end{figure}

Without applying quality criteria, the zenith angle distribution of the 
reconstructed muons is strongly smeared. 
Therefore we have calculated the elements of the parent angular distribution  
$N_{\mu}(\theta_{\mu})$ from the reconstructed distribution 
$N_{\mu}(\theta_{rec})$ using a standard regularized deconvolution
procedure \cite{blobel1}.
The flux $I(\theta)$ can be transformed into a
vertical flux $I(\theta=0,h)$, where $h$ is the ice thickness in
mwe (meter water equivalent) seen under angle $\theta$:
\begin{equation}\label{vertflux}
I(\theta=0,h)=I(\theta) \cdot \cos{\theta} \cdot c_{corr}
\end{equation}
The $\cos{\theta}$-conversion corrects the
sec~$\theta$ behaviour of the muon flux, valid
for angles up to 60$^o$.
The term  $c_{corr}$ is taken from \cite{Lipari} and corrects
for larger angles. It varies between 0.8 and 1.0 for the angular 
and energy ranges considered here.
The vertical intensities obtained in this way are plotted in
fig.\ref{ABD}. The results are in agreement with the depth-intensity 
published by DUMAND \cite{dumand}, Baikal \cite{baikal}, and the prediction 
given by Bugaev et al. \cite{bugaev}.

\subsection{Search for Upward Going Muons}
\noindent
AMANDA-B4 was not intended to be a 
full-fledged neutrino detector, 
but instead a device which demonstrates the 
feasibiliy of muon track
reconstruction in Antarctic ice.  
The  limited number of optical modules and
the small lever arms in all but the vertical direction 
complicate the rejection of fake events. 
In this section we demonstrate
that in spite of that
the separation of a few upward
muon candidates was possible.

\begin{figure}[htbp]
\centering
\mbox{\epsfig{file=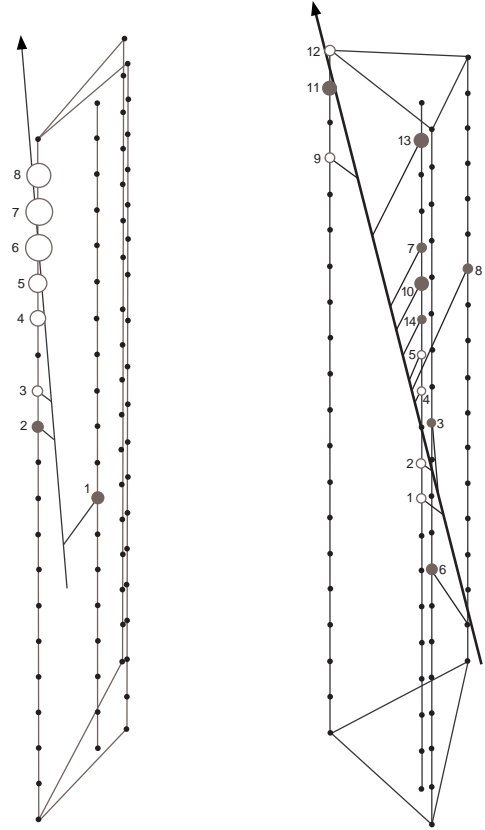,height=11cm}}
\caption[2]{\small
The two experimental events reconstructed as upward muons.
{\it left:} ID 8427907, {\it right:} ID 4706870. 
The line with an arrow symbolizes the fitted muon track, the lines
from this track to the OMs  indicates light pathes. 
The amplitudes are proportional to the size of the OMs.
The numbering of the
OMs refers to the time order in which they are hit.
}
\label{twoevents}
\end{figure}

Two full, but independent analysis were performed with the
experimental data set of 1996. 
In the first analysis, a 
fast pre-filter reduced the background Monte Carlo sample 
to 5\%, whereas
50\% of simulated upgoing events survived \cite{bouchta}.
Full reconstruction and application of the  criteria
\vspace{-3mm}
\begin{enumerate}
\item Hits on $\ge$ 2 strings
\vspace{-3mm}
\item  Reconstructed zenith angle $\theta >$ 90$^o$
\vspace{-3mm}
\item $\log({\cal L}_{time})/N_{hit} > -6$
\vspace{-3mm}
\item $\vec{v_z} \ge$ 0.15 m/nsec
\vspace{-3mm}
\item  $N_{dir}(15) \ge$6
\end{enumerate}
\vspace{-2mm}
reduces the experimental sample to 2 events,  
in agreement with the Monte Carlo expectation of 2.8 
for atmospheric neutrinos. The two events are
shown in fig.\ref{twoevents}.

For the second analysis, 
all events have been reconstructed, with different, independent
likelihood parametrization and minimisation programs,
and then reduced
by the subsequent application of the following criteria:
\vspace{-2mm}
\begin{enumerate}
\item zenith angle of the line approximation and of the 
   reconstruction $\theta > 120^o$,
\vspace{-3mm}
\item speed of the line approximation $ 0.15 <  |\vec{v}| < 1$ m/nsec,
\vspace{-3mm}
\item $\log({\cal L}_{time})/(N_{hit}-5) > -10$,
\vspace{-3mm}
\item ${\cal L}_{hit}/(N_{hit}-5) > -8$,
\vspace{-3mm}
\item $N_{dir}(25) \ge 5 $,
\vspace{-3mm}
\item $N_{dir}(75) \ge 9 $,
\vspace{-3mm}
\item $L_{dir}(25) > 200$\,m,
\vspace{-3mm}
\item $ |z_{COG}| < 90$m (absolute value of the vertical coordinate 
of the center of gravity given by hit OMs).
\end{enumerate}
\vspace{-2mm}
These cuts reduced the experimental data sample to 3 events.
The passing rate for Monte Carlo upward moving muons from atmospheric
neutrinos is 1.3\%, giving an expectation of 4 events. 
Two of the three experimental events were also identified in the
previously described search,  one, however, did not pass the
the cut on direct hits ($N_{dir}$=5 instead of 6).

In order to check how well the parameter distributions of the
events agree with what one expects for atmospheric neutrino
interactions, and how well they are separated from the
rest of the experimental data, we relaxed two cuts at a time 
(retaining the rest) and  inspected the distribution in the
two "free" variables.
\vspace{4mm}
\begin{figure}[htbp]
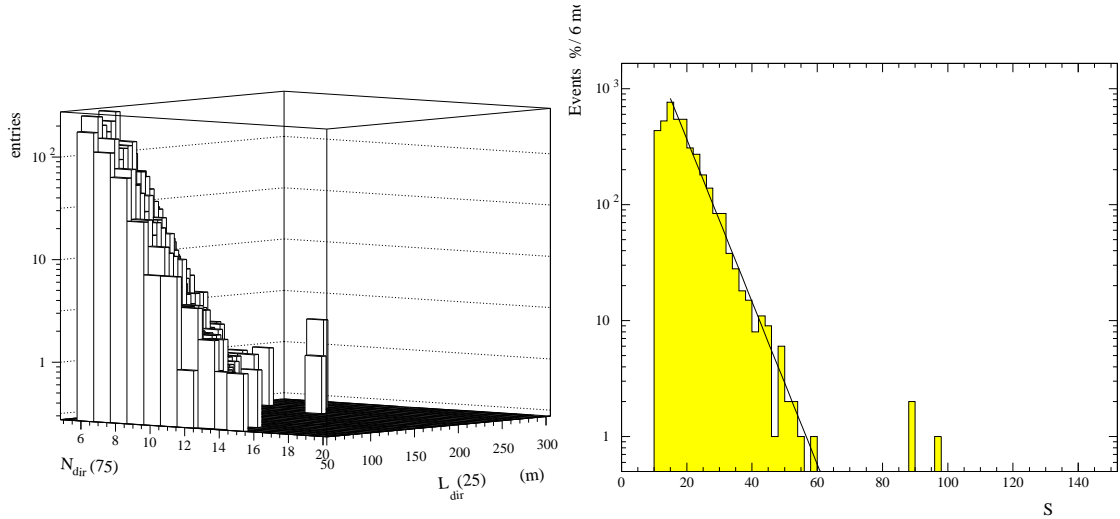

\centering
\mbox{\epsfig{file=96_sepa_3d.epsi,width=.48\textwidth}}
\mbox{\epsfig{file=96_sepa_1d.epsi,width=.48\textwidth}}
\vspace{-5mm}
\caption[2]{\small
After application of cuts with the exception of 
cuts 6 and 7:
{\it left} --  distribution in parameters $L_{dir}$(25)
vs. $N_{dir}$(75),  
{\it right:} distribution in the "combined" parameter
($(N_{dir}$(75)-2) $\cdot$ $L_{dir}$(25)/20\,\mbox{m}.)
\label{sepa}
}
\end{figure}

Fig.~\ref{sepa} shows, as an example, 
the distribution in $L_{dir}$(25)
and $N_{dir}$(75). The three events
passing {\it all} cuts are  separated from
the bulk of the data. At the bottom of fig.~\ref{sepa},
the data are plotted
versus a combined parameter, 
$  S = (N_{dir}$(75)-2) $\cdot L_{dir}$(25)/20.
In this parameter, the  data exhibit
a nearly exponential decrease. Assuming the decrease of the
background dominated events to continue at higher $ S $ values,
one can calculate the probability that the separated events are
fake events. The probability to observe one event
at $S \ge 70$ is 15\%, the probability to
observe 3 events is only $6 \cdot 10^{-4}$.

We conclude that tracks reconstructed as up-going
are found at a rate consistent with that
expected for  atmospheric neutrinos. The three events
found in the second analysis
are well separated from background. In a limited
angular interval,
even with a detector as small as AMANDA-B4, neutrino
candidates can be separated. 

\vspace{-2mm}

\section{Preliminary results from AMANDA-B10}

\vspace{-7mm}

\subsection{Separation of atmospheric neutrino events}

\begin{figure}[htbp]
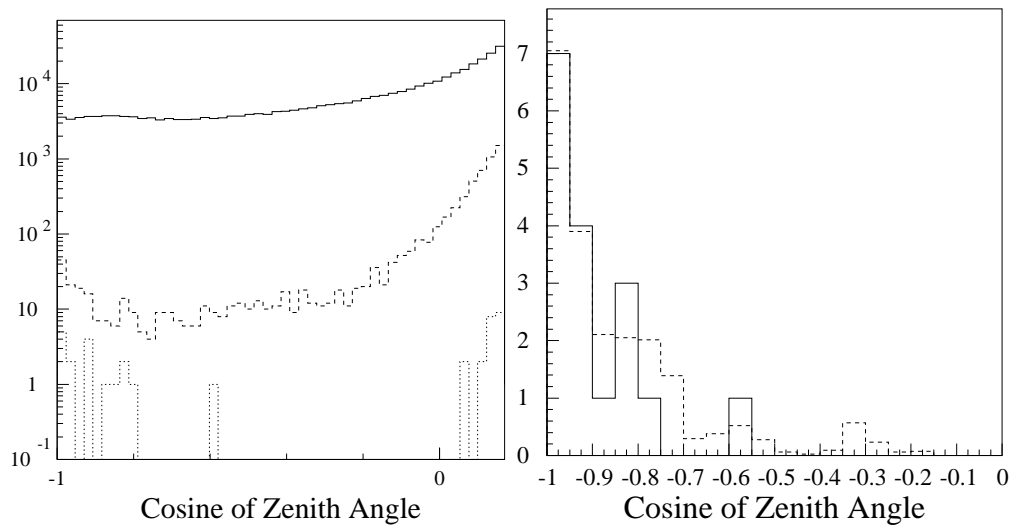

\centering
\mbox{\epsfig{file=Alb2.epsi,width=.43\textwidth}}
\mbox{\epsfig{file=Alb1.epsi,width=.43\textwidth}}
\caption[2]{\small
Left: The reconstructed zenith angles of 113 days of 
AMANDA-B10 data after quality cut levels 2 to 4
(from top to bottom). Right: Zenith angle
distribution of neutrino candidates (solid line) and
of Monte Carlo simulated atmospheric neutrino events
normalized to 85 days effective live time (dotted line)
at cut level 4. 
}
\label{Alb1}
\end{figure}

\noindent
We have performed a first analysis of data taken during a period of
113 days of the first year of operation (1997) of AMANDA-B10 
\cite{Karle,Hill}.
The corresponding effective live time of the detector is about
85 days. The total experimental sample consists of
$4.9 \cdot 10^8$ events. The events were filtered and reconstructed.
After that, a set of quality/upward-muon criteria has been applied. 
In this first approach, cuts have been applied only to
a subset of the parameters used for the B4 analysis. The
cuts were grouped into four levels of subsequently increasing
tightness. 17 events passed all four cuts, compared to 21.1
events predicted by Monte Carlo. Fig.\ref{Alb1} shows the distribution in
$\cos \theta$  after cut 2, 3 and 4, respectively. We note
that a fine-tuned analysis is expected to yield
2-3 times more events! 

\bigskip

\begin{figure}[htbp]
\centering
\mbox{\epsfig{file=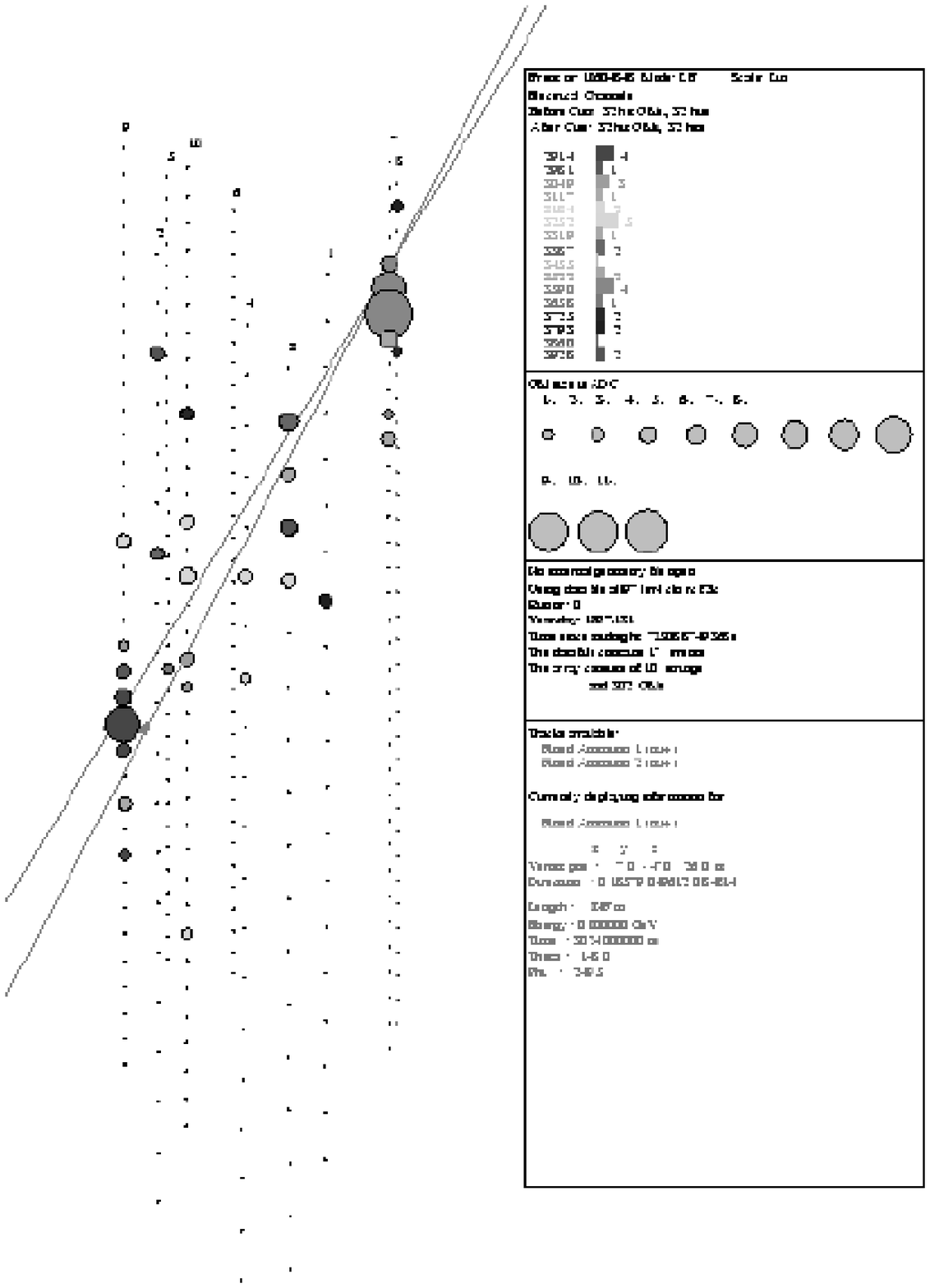,height=14.5cm}}
\caption[2]{\small
An upgoing muon candidate in Amanda-B10,
with 32 hits on 9 strings. The amplitudes are proportional
to the diameter of the full circles. In a colored view, their
color indicates the time relative to the trigger time. 
The 2 lines show results of 
line approximation and final fit.}
\label{Alb2}
\end{figure}

The 17 upward
muon candidates are seen on the left side (the tail of
a few additional events on the right side appears since
the angular cut were
at 80 degrees instead of 90 degrees). Fig.\ref{Alb2} shows
the zenith angle distribution of the 17 neutrino candidates and
of Monte Carlo simulated atmospheric neutrinos.  Within the
limited statistics, one observes satisfying agreement.
Fig.\ref{Alb2} displays one neutrino event. Compared
to fig.\ref{twoevents}, it illustrates the significant gain in
complexity and information obtained by moving from
4 strings to the 10-string array.


\newpage

\subsection{Search for relativistic magnetic monopoles}

\noindent
Magnetic monopoles   with unit magnetic Dirac charge 
$g = 137/2 \cdot e$ and
velocities above the Cherenkov threshold in water ($\beta > 0.75$)
would emit a huge amount of light.
Its Cherenkov radiation exceeds that of a bare relativistic
muon by a factor $(g \cdot n/e)^2 = 8300$, with $n$=1.33 being the
index of refraction for water \cite{Mon,Janmon}. We therefore searched
for events with high hit multiplicity \cite{Peter}. We analyzed 45
days of effective live time of the 1997 B10 data,
yielding $1.8 \cdot 10^6$ events with mulitplicity larger
than 75. 
In order not to be
dominated by brems-showers along downward muons, we applied
cuts on the zenith angle $\Theta$ given by the line fit.
Obscure time patterns where rejected by cuts on the
fitted velocity $v$. To reject
high multiplicity events due to cross talk along the cables, special
cuts on time differences of hits along
one string have been applied.

\begin{figure}[htbp]
\centering
\mbox{\epsfig{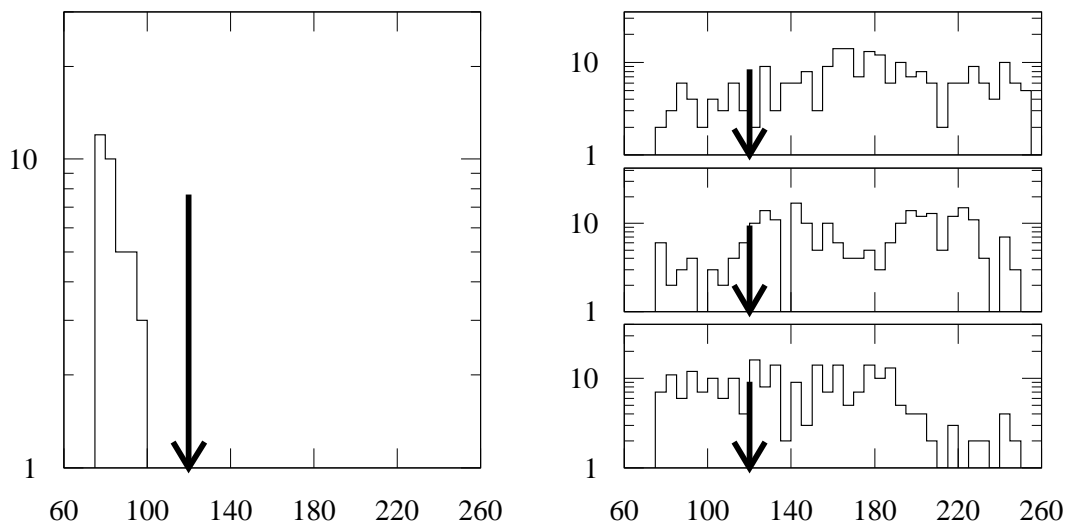}}
\caption[2]{\small
Hit multiplicity of events after application of cuts (see text).
Left: experimental data. Right: distributions for 
monopoles with $\beta$= 1.0, 0.9, 0.8 (from top to bottom). The arrow
indicates the final multiplicity cut.
}
\label{Peter1}
\end{figure}

Monopoles with velocities $\beta$ =0.8, 0.9 and 1.0
have been simulated and tracked through the detector.
Fig.\ref{Peter1} shows the multiplicity of experimental events
after application of all cleaning criteria (left) and that of
simulated magnetic monopoles of different velocity after the same
criteria. The arrow indicates a cut at multiplicity 120
(clearly above the maximum of 100 hits observed).

\begin{figure}[htbp]
\centering
\mbox{\epsfig{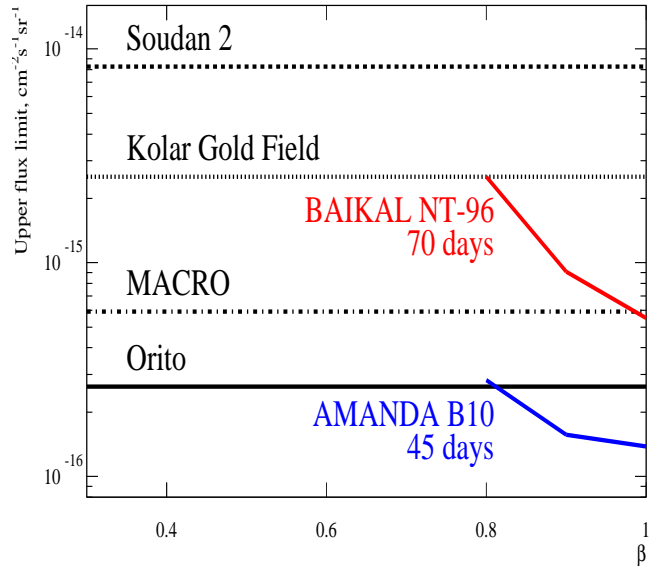}}
\caption[2]{\small
Upper flux limit (90\%C.L. for various experiments)
}
\label{Peter2}
\end{figure}

With the
acceptance for monopoles after all cuts, including the
$N_{hit} > 120$ cut, being 
20.5 (16.0, 9.8)$\cdot 10^9$cm$^2$sr for $\beta$ = 1.0 (0.9,0.8),
and no experimental event with
$N_{hit} > 100$, 
the limit shown in fig.\ref{Peter2}
is obtained. This limit applies to monopoles with
masses larger than $10^{10}-10^{11}$ GeV since lighter
monpoles would have been stopped in the Earth.

\subsection{Other directions of analysis}

\noindent
Results on atmospheric neutrinos and on magnetic
monopoles are just examples for a broad front of analyses 
underway. We mention the following, most of them being reported
in contributions to the 26th ICRC:

\begin{itemize}
\vspace{-2mm}
\item Search for point sources of neutrinos. With an area
of a few $10^3$ m$^2$ for TeV neutrinos, AMANDA-B10 is
the most sensitive high energy neutrino telescope\cite{John}.
\vspace{-5mm}
\item Search for an excess of events over 
atmospheric neutrinos due to WIMP annihilation in the
center of the Earth \cite{Eva}.
\vspace{-2mm}
\item Search for high energy cascades, similar to
our early analysis of AMANDA-A data \cite{Rodin}
and in analogy to the
Baikal analysis presented at this Workshop \cite{Jan}.
\vspace{-2mm}
\item Search for neutrino events in coincidence with GRB coincidences.
Since only short time windows bracketing the GRB are scanned, the
background rejection criteria can be loosened considerably,
resulting in a much higher effective area than in
the standard point search analysis \cite{Ryan}.
\vspace{-2mm}
\item Search for counting rate excesses due to a supernova
explosion \cite{Ralf}. 
A future alert algorithm will enable AMANDA to contribute
to a worldwide alert network.
\vspace{-2mm}
\item Investigation of seasonal variations of trigger rates which
are closely correlated to temperature and pressure of the
atmosphere above the South Pole \cite{Adam}.
\vspace{-2mm}
\end{itemize}

\section{Future development}

\noindent
In the season 1999/2000, we plan to deploy
six additional strings 
which complete the 30,000 m$^2$ telescope AMANDA-II. 
The new strings
will also be used to test a variety of new techniques. Part of the
OMs will contain $10^{\prime \prime}$ PMTs with about 50\%
better light collection. Possibly, wavelength shifters will be
applied, which increase the sensitivity in the UV and
might give another factor of 30-40\% in light collection. 
The analog transmission
of optical signals will be improved, making use of better electronics
schemes and of laser diodes instead of LEDs. Another part of the
R\&D effort is the construction and deployment of a string equipped
with digital optical modules in order to investigate
waveform digitization at depth \cite{DOM}. 
This waveform is then transmitted
via a serial link to the surface. The method is challenging
since all OMs have to be
synchronized on a nanosecond time scale and a complicated
communication has to be performed over 2 km electrical cable..

The long-term goal of the collaboration is a detector of the
scale of a cube kilometer\cite{Francis}. 
This is the order of magnitude
suggested by many models of neutrino production in AGN,
in the center of the Galaxy, in the center of the Earth
(due to WIMP annihilation) and in young supernovae \cite{GHS}. 
A straw-man design calls for a total of about 5000 PMTs at 80 strings, 
horizontally spaced by 80-100 meters. The strings would be
instrumented between 1.4 and 2.4~km. The approximate
cost for the detector can be derived from the
estimated cost of 6-8 k\$ per channel 
(including cables and electronics), and amounts to 
about 35 M\$.
Construction of ICECUBE will be staged over five to six
deployments (possibly 2002/03 to 2007/08.)

\section{Acknowledgments}

\noindent
{\footnotesize
This research was supported by 
the U.S. National Science Foundation, Office of Polar Programs
and Physics Division,
the University of Wisconsin Alumni Research Foundation,
the U.S. Department of Energy,
the U.S. National Energy Research Scientific
Computing Center, 
the Swedish Natural Science Research Council,
the Swedish Polar Research Secretariat,
the Knut and Alice Wallenberg Foundation, Sweden,
and the Federal Ministery for Education and Research, Germany.}

\end{document}